\documentclass[11pt,aps,amsmath,prc,nofootinbib,superscriptaddress]{revtex4-2}

\usepackage{graphicx} 
\usepackage{bm}
\usepackage{slashed} 
\usepackage{hyperref}

\newcommand{\mc}[1]{\mathcal{#1}}
\newcommand{\sgn}{\mathrm{sgn}}

\begin{document}
\title{Searching for missing direct photons in heavy-ion collisions with P and CP violation}

\author{Jonathan D. Kroth}
\email{jdkroth@iastate.edu}
\affiliation{Department of Physics and Astronomy, Iowa State University, Ames, Iowa 50011, USA}

\author{Kirill Tuchin}
\email{tuchink@gmail.com}
\affiliation{Department of Physics and Astronomy, Iowa State University, Ames, Iowa 50011, USA}

\date{\today}

\begin{abstract}

We compute synchrotron radiation from a plasma in which $P$- and $CP$-violating parameters, a chiral chemical potential and a chiral gradient, couple to fermions. To do this, we compute exact wavefunctions for the fermions in the presence of these parameters and an external constant magnetic field. We find that these parameters increase the synchrotron radiation emitted by the fermions while also decreasing the traditionally large synchrotron radiation elliptic flow coefficient $v_2$. We apply these results to the quark-gluon plasma, where just such a contribution could provide a solution to the missing direct photons puzzle. We also use our wavefunctions to give a derivation of the chiral magnetic effect.

\end{abstract}

\maketitle

\section{Introduction}

The traditional mechanism of direct photon production in relativistic heavy-ion collisions relies on conventional perturbation theory to describe short-distance processes and hydrodynamic approximation to account for the collective behavior of the quark-gluon plasma (QGP). This approach has proven fairly successful in providing a phenomenological framework \cite{Arnold:2001ms,Ghiglieri:2013gia,Traxler:1994hy,Paquet:2015lta,vanHees:2011vb,Shen:2013vja,Wu:2025iix,Baier:1991em,Kapusta:1991qp,Hung:1996mq,Steele:1996su,Dusling:2009ej,Lee:1998nz,Aurenche:2000gf,Peitzmann:2001mz,Turbide:2003si,Bratkovskaya:2008iq,Vitev:2008vk,vanHees:2011vb,Linnyk:2015tha}. However, it faces challenges in simultaneously describing the soft part of the photon spectrum and its variation with the azimuthal angle around the collision axis \cite{PHENIX:2014nkk,PHENIX:2022rsx,PHENIX:2015igl,PHENIX:2025ejr}. The state-of-the-art theoretical predictions use a combination of initial state glasma, hydrodynamic, and hadron resonance gas models to predict the photon spectrum, but fails to predict a sufficient number of photons and gives too mild a variation with the azimuthal angle, as quantified by the elliptic flow \cite{Gale:2021emg}.

Several solutions have been proposed to solve this problem \cite{Chiu:2012ij,McLerran:2014hza,Basar:2012bp,Tuchin:2012mf,Yee:2013qma,Fukushima:2012fg,Mamo:2013jda,Mamo:2015xkw,Tuchin:2025bll,Tuchin:2025stl,Buzzegoli:2023yut,Buzzegoli:2022dhw,Buzzegoli:2025qfl}. Among these, one of the most promising approaches involves incorporating the synchrotron photons emitted by the QGP due to the presence of a strong magnetic field. This synchrotron radiation not only provides additional photons, but also exhibits a highly anisotropic distribution. As a result, it significantly enhances the elliptic flow, as most of the synchrotron radiation is emitted perpendicular to the magnetic field.  In fact, the elliptic flow of photons generated by the strong magnetic field exceeds the experimental data. The situation is further complicated in the presence of rotation, which tends to amplify the effect of the magnetic field, as we have recently demonstrated in \cite{Buzzegoli:2025qfl,Kroth:2024mfd,Buzzegoli:2023vne,Buzzegoli:2022dhw}.

In this paper, we argue that the elliptic flow $v_2$ strongly depends on the chiral properties of the QGP. Specifically, it depends on the transport coefficients associated with the chiral magnetic and anomalous Hall currents. These currents are directly associated with the chiral anomaly and, as such, violate $P$- and $CP$-symmetries. In particular, we find that non-zero values for either of these parameters significantly reduces $v_2$. The reason that the chiral magnetic current reduces the elliptic flow may be that the current flows in the direction of the magnetic field and the relativistic quarks emit photons mostly in the forward direction. In contrast, synchrotron radiation is emitted mostly perpendicular to the magnetic field. A different mechanism of photon production via the chiral anomaly in the presence of a magnetic field, but ignoring synchrotron radiation, was previously considered in \cite{Kharzeev:2013wra,Tuchin:2019jxd,Mamo:2015xkw,Fukushima:2012fg,Yee:2013qma,Jia:2022awu,Jia:2024ctx}.

Our objective is to develop a consistent theoretical framework for computing photon radiation in the presence of anomalous currents. A powerful approach to accomplish this goal is to employ an effective field theory which adds to the standard electromagnetic interaction the following term \cite{Colladay:1996iz,Kostelecky:2002ue,Kharzeev:2007jp,Sukhachov:2018uuz,Sheng:2017lfu}:
\begin{equation}\label{eq:DA}
    \mathcal{L}_\text{anom}= \partial_\mu \theta j^{\mu 5} = b_\mu j^{\mu 5}\,,
\end{equation}
which couples the chiral vector current $j^{\mu 5}=\bar \psi \gamma^\mu\gamma^5\psi$ to the background pseudoscalar field $\theta$. The equations of motion depend on the vector field $b_\mu=\partial_\mu \theta$,\footnote{In this paper it is convenient to define $b_\mu= \partial_\mu \theta$, consistent with \cite{Colladay:1996iz,AlanKostelecky:2024psy,Kostelecky:2002ue}. Another definition $b_\mu= c_A\partial_\mu \theta$ is also often used in the literature,  e.g.\ \cite{Qiu:2016hzd,Hansen:2022nbs,Hansen:2023wzp,Hansen:2025gzt,Hansen:2024kvc,Tuchin:2025bll}.} whose time component has the physical meaning of a chiral chemical potential $\mu_5$, while the spatial components are most familiar as the displacement in momentum space of the Weyl nodes of a semimetal.  The anomalous currents associated with $b_\mu$ are the chiral magnetic current $\bm j = c_Ab_0\bm B$ and the anomalous Hall current $\bm j = c_A\bm b\times \bm E$. Throughout this paper, $b_\mu$ is assumed to be constant.

Alternatively, the same physics can be described by coupling the field invariant $F_{\mu\nu} \tilde{F}^{\mu\nu}$ to $\theta$. We have pursued this method in our previous publications \cite{Tuchin:2014hza,Tuchin:2016tks,Tuchin:2018sqe,Tuchin:2019jxd}. 
The two effective theories are related to each other in the chiral limit by a chiral transformation.   Therefore, while each approach offers a different vantage point, in the chiral limit, the final results must remain independent of the choice of the effective theory. However, establishing such equivalence in practice may be  challenging. 

This paper employs the approach based on the effective term given by Eq. (\ref{eq:DA}).  The resulting equation of motion, including coupling to a background magnetic field, is given by Eq. (\ref{eq:Dirac}). Our goal is to solve this equation and utilize its solutions to investigate the photon production by the QGP. 
 
Recently, Shovkovy and Wang conducted a similar calculation \cite{Wang:2024gnh}. They employed a relationship between the photon production rate and the fermion propagator in a magnetic field at a finite chiral chemical potential $\mu_5$, which is identified with $b_0$. Unlike \cite{Wang:2024gnh}, we do not restrict ourselves to the chiral limit and also take into account $b_3$, associated with the longitudinal component of the anomalous Hall current.

The paper is structured as follows: in Section \ref{sec:Spinors}, we derive solutions to the modified Dirac equation given by (\ref{eq:Dirac}) in a constant magnetic field. These spinors, expressed by (\ref{eq:DiracSpinorFinal}), constitute our main theoretical result. We turn to phenomenology in Section \ref{sec:Rad}, where we use these spinors to compute the synchrotron radiation emitted by these fermions. We complete this calculation numerically in Section \ref{sec:Num}. We conclude in Section \ref{sec:Conc}. As another application of our wavefunctions, we compute the chiral magnetic current using them in Appendix \ref{sec:CME}.

We employ the natural units $\hbar=c=k_B=1$ and use the following notations: $p^\mu = (E, \bm p)$, $k^\mu=(\omega,\bm k)$, $b^\mu=(b_0,\bm b)$, with $b_0=\mu_5$ being the chiral chemical potential. We refer to $\bm b$ as the chiral gradient.

\section{Dirac Equation with Chiral Couplings}
\label{sec:Spinors}

We wish to solve the Dirac equation
\begin{equation}
    (i \slashed{D} - \gamma^5 \slashed{b} - m) \psi = 0\,,
    \label{eq:Dirac}
\end{equation}
where $D$ is the covariant derivative, for a particle of charge $q$ and mass $m$ in the presence of a constant magnetic field $\bm{B}$ which points along the 3-axis. The field $b_\mu$ is assumed to be constant, with only two non-vanishing components: $b_0$ and $b_3$.

We use the general gauge choice
\begin{equation}
    A_\mu = (0, aBy, -(1-a)Bx, 0)
\end{equation}
with $a \in [0,1]$. $a = \tfrac{1}{2}$ corresponds to the symmetric Landau gauge, which we will use explicitly later when it becomes useful to choose a gauge. 

\subsection{Determination of Two-Component Spinors}
To solve Eq. (\ref{eq:Dirac}), we write
\begin{equation}
	\psi(x) = e^{-iEt+ip_3 z} \begin{pmatrix} \chi_-(\bm{x}_\perp) \\ \chi_+(\bm{x}_\perp)
    \end{pmatrix} .
\end{equation}
with $\chi_\sigma$ two-component spinors. Throughout this paper, we allow $E$ to be negative. For many cases, this merely means an antiparticle. However, the dispersion relation we will derive will not in general give antiparticles with the same (absolute value of) energy as particles. We use $\sigma = \pm 1$ throughout this paper to denote chirality. In the Weyl representation, the Dirac equation (\ref{eq:Dirac}) gives
\begin{equation}
	\begin{pmatrix}
		E - b_3 - \sigma (p_3 - b_0) & \sigma D_{-}^\perp \\
		\sigma D_{+}^\perp & E + b_3 + \sigma (p_3 + b_0)
	\end{pmatrix} =: \Pi_\sigma \chi_\sigma = m \chi_{-\sigma}
	\label{eq:DiracFirst}
\end{equation}
Note as well that $b_3 = \bm{b} \cdot \hat{z}$ is the spatial component of $b^\mu$. We have defined, for $\tau = \pm 1$,
\begin{equation}
	D_{\tau}^\perp = (iD_1 - \tau D_2) .
\end{equation}
We can decouple these first-order equations by making them second order:
\begin{equation}
	\begin{pmatrix}
		(E - b_3)^2 - (p_3-b_0)^2 - D_-^\perp D_+^\perp & -2 D_-^\perp (b_0 + \sigma b_3) \\
		-2 D_+^\perp (b_0 - \sigma b_3) & (E+b_3)^2 - (p_3+b_0)^2 - D_+^\perp D_-^\perp
	\end{pmatrix} \chi_\sigma = \Pi_{-\sigma} \Pi_\sigma \chi_\sigma = m^2 \chi_\sigma .
    \label{eq:DiracSquared}
\end{equation}

Now, it is known that when $b_0=b_3=0$, the eigenfunctions $\psi(x)$ are simple harmonic oscillator wavefunctions; in that case, the operators $D_\pm^\perp$ serve as the raising and lowering operators. Since $\slashed{b} \gamma^5$ commutes with $\slashed{D}^\perp$, these simple harmonic oscillator states will continue to be useful as spatial eigenfunctions. Notice that, had we allowed $b_1$ or $b_2$ to be non-zero (equivalent by a change of coordinates and a gauge transformation), this would not be the case, and our analysis would be much more difficult.

To this end, choose an orthonormal basis $\{\phi_n(\bm{x}_\perp)\}_n$ of spinor simple harmonic oscillator eigenfunctions appropriate to the choice of gauge.\footnote{The eigenfunctions $\phi_n$ are two-component spinors with two indices, $n$ and another appropriate to the gauge choice. For $a=\tfrac{1}{2}$, it is the azimuthal angular momentum $j$, while for $a=1$, it is $p_1$. We treat these additional points in Appendix \ref{app:LandauLevels}, as they do not affect the derivation of the wavefunctions.} A straightforward calculation shows
\begin{equation}
	D_\tau^\perp \phi_n = - \sqrt{2|qB| \left(n + \frac{1 - \tau \beta}{2}\right)} \phi_{n - \tau \beta} , \qquad D_\tau^\perp D_{-\tau}^\perp \phi_n = 2 |qB| \left(n + \frac{1 + \tau \beta}{2}\right) \phi_n ,
	\label{eq:RaiseLower}
\end{equation}
where $\beta = \sgn(qB)$. Write $\chi_{\beta\sigma}$ in terms of these eigenfunctions,
\begin{equation}
	\chi_{\beta \sigma} = \sum_{n=0}^\infty \begin{pmatrix} c_n^{\beta \sigma} \phi_n \\ d_{n}^{\beta \sigma} \phi_n \end{pmatrix} ,
\end{equation}
where we have introduced a $\beta$ subscript due to its appearance in Eq. (\ref{eq:RaiseLower}). Using the orthonormality of the $\phi_n$, one finds
\begin{equation}
	0 = \begin{pmatrix}
		E^2 - \Lambda_n^2 - 2X & 2(b_0 + \sigma b_3) \sqrt{2|qB|n} \\
		2 (b_0 - \sigma b_3) \sqrt{2|qB|n} & E^2 - \Lambda_n^2 + 2X
	\end{pmatrix}
	\begin{pmatrix} c_n^{\sigma} \\ d_{n}^{\sigma} \end{pmatrix} ,
	\label{eq:CoeffMatrixEqn}
\end{equation}
where we have defined
\begin{equation}
	\Lambda_n^2 = p_3^2 + 2n|qB| + b^2 + m^2 \quad \text{and} \quad X = E b_3 - p_3 b_0 .
    \label{eq:LambdaX}
\end{equation}
We have removed the index $\beta$ by defining $c_n^\sigma = c_n^{+ \sigma} = c_{n-1}^{- \sigma}$ and $d_n^\sigma = d_{n-1}^{+ \sigma} = d_{n}^{- \sigma}$, for $n > 0$. That is, the equations determining $c_n^{\beta\sigma}$ and $d_n^{\beta\sigma}$ decouple into pairs of equations for $c_n^{\beta\sigma}$ and $d_{n-\beta}^{\beta\sigma}$. Had we allowed $b_1$ or $b_2$ to be non-zero, this would not have occurred --- this is the difficulty mentioned above --- and signals that a different spatial basis should be used. So there are eigenspinors
\begin{equation}
	\chi_{\beta\sigma}^{(n)} = \begin{pmatrix} c_n^\sigma \phi_{n-\left(\frac{1-\beta}{2}\right)} \\ d_{n}^\sigma \phi_{n-\left(\frac{1+\beta}{2}\right)} \end{pmatrix}
\end{equation}
It then remains to solve Eq. (\ref{eq:CoeffMatrixEqn}) for $n > 0$ and to analyze the special case of $n=0$.

\subsubsection{Special Case: $n=0$}
In the case $n = 0$, we retain the additional index $\beta$. The same analysis leading to Eq. (\ref{eq:CoeffMatrixEqn}) leads to
\begin{gather}
    (E^2 - \Lambda_0^2 - 2X) c_0^{+ \sigma} = 0 \\
    (E^2 - \Lambda_0^2 + 2X) d_0^{- \sigma} = 0 .
\end{gather}
All other terms trivially vanish since $c_{-1}^{\beta\sigma} = d_{-1}^{\beta\sigma} = 0$, by definition. Then, for a non-vanishing $n=0$ spinor, $c_0^{+ \sigma} = d_0^{- \sigma} = 1$, and we get the dispersion
\begin{equation}
    (E - \beta b_3)^2 - (p - \beta b_0)^2 - m^2 = 0 .
    \label{eq:n0Dispersion}
\end{equation}
The $n=0$ spinor is then
\begin{equation}
    \chi_{\beta\sigma}^{(0)} = \begin{pmatrix} \frac{1+\beta}{2} \\ \frac{1-\beta}{2} \end{pmatrix} \phi_0 .
\end{equation}

\subsubsection{General Case: $n>0$}
For $n \geq 1$, Eq. (\ref{eq:CoeffMatrixEqn}) gives the dispersion relation
\begin{equation}
    (E^2 - \Lambda_n^2 - 2X) (E^2 - \Lambda_n^2 + 2X) = (E^2 - \Lambda_n^2)^2 - 4X^2 = 4b^2 2n|qB| .
    \label{eq:Dispersion1}
\end{equation}
This is a quartic in $E$ whose solution cannot be succinctly written. For later use, we write this in another form,
\begin{equation}
    ((\tilde{p} + b)^2 - m^2)((\tilde{p} - b)^2 - m^2) = -4 b^2 m^2 ,
    \label{eq:Dispersion2}
\end{equation}
with the shorthand $\tilde{p}^\mu = (E,0,\sqrt{2n|qB|},p_3)$.

Solving Eq. (\ref{eq:CoeffMatrixEqn}) and normalizing $\chi_{\beta\sigma}^{(n)}$ gives
\begin{gather}
    |c_n^{\sigma}|^2 = \frac{1}{2} \frac{(b_0 + \sigma b_3)(E^2 - \Lambda_n^2 + 2X)}{b_0 (E^2 - \Lambda_n^2) + 2 \sigma b_3 X} ,
    \label{eq:Csq} \\
    |d_{n}^{\sigma}|^2 = \frac{1}{2} \frac{(b_0 - \sigma b_3)(E^2 - \Lambda_n^2 - 2X)}{b_0 (E^2 - \Lambda_n^2) + 2 \sigma b_3 X} .
    \label{eq:Dsq}
\end{gather}
Appealing to Eq. (\ref{eq:CoeffMatrixEqn}) again, we find the relative sign between $c_n^\sigma$ and $d_n^\sigma$ to be
\begin{equation}
    s^\sigma := -\sgn\left(\frac{E^2-\Lambda_n^2-2 \tau X}{b_0 + \sigma \tau b_3}\right) ,
    \label{eq:Ssigdef}
\end{equation}
for either choice of $\tau = \pm1$. We will insert this sign by hand by choosing $c_n^\sigma$ and $d_n^\sigma$ to be positive and putting $s^\sigma$ in the second component of $\chi_{\beta\sigma}^{(n)}$, i.e.
\begin{equation}
    \chi_{\beta \sigma}^{(n)} = \begin{pmatrix} c_n^{\sigma} \phi_{n+\bar{\beta}} \\ s^\sigma d_{n}^{\sigma} \phi_{n+\overline{-\beta}} \end{pmatrix} .
\end{equation}
The two signs are also related:
\begin{equation}
    s^\sigma = -\sgn(b^2) \sgn\left(\frac{E^2-\Lambda_n^2-2X}{b_0 - \sigma b_3}\right) = \sgn(b^2) s^{-\sigma} .
\end{equation}
Notice that $s^\sigma$ is not well-defined in the limit $b_0,b_3 \to 0$. In that limit, the matrix equation for the two-component spinors becomes diagonal, signaling that we require more constraints to specify $c_n^\sigma$ and $d_n^\sigma$ uniquely. This is usually done by also requiring $\psi$ to be an eigenspinor of a spin-like operator.

\subsection{Determination of Dirac spinors}
Now we will glue together the $\chi_{\beta\sigma}^{(n)}$ to form solutions to the Dirac equation. Write
\begin{equation}
    \psi^{(n)}_\beta = \begin{pmatrix} f_n^\beta \chi_{\beta -}^{(n)} \\ g_n^\beta \chi_{\beta +}^{(n)} \end{pmatrix} .
\end{equation}
The Dirac equation then gives an equation for the coefficients,
\begin{equation}
    \begin{pmatrix}
        E-b_3 - \sigma(p_3-b_0) & -\sigma \sqrt{2n|qB|} \\
        -\sigma \sqrt{2n|qB|} & E+b_3 + \sigma(p_3+b_0)
    \end{pmatrix}
    \begin{pmatrix} c_n^\sigma \\ d_n^\sigma \end{pmatrix}
    = m \left(\frac{f_n^\beta}{g_n^\beta}\right)^\sigma
    \begin{pmatrix} c_n^{-\sigma} \\ d_n^{-\sigma} \end{pmatrix}
\end{equation}
where on the right-hand side $\sigma$ is an exponent on the fraction rather than a label. At this point, it is evident that $f_n^\beta$ and $g_n^\beta$ do not depend on $\beta$ for $n>0$, so we will drop the associated superscripts.

\subsubsection{Special Case: $n=0$}
We first handle the $n=0$ case. Since one of $c_0^{\beta\sigma}$ and $d_0^{\beta\sigma}$ is 0, we get one equation to solve,
\begin{equation}
    \left(\frac{f_0^\beta}{g_0^\beta}\right)^\sigma = \frac{E-\beta b_3 - \beta \sigma(p_3 - \beta b_0)}{m} .
\end{equation}
Normalizing the spinor gives
\begin{gather}
    |f_0^\beta|^2 = \frac{E - \beta b_3 - \beta (p_3 - \beta b_0)}{2(E - \beta b_3)} , \\
    |g_0^\beta|^2 = \frac{E - \beta b_3 + \beta (p_3 - \beta b_0)}{2(E - \beta b_3)} .
\end{gather}
An analysis of the $n=0$ dispersion relation shows that
\begin{equation}
    \sgn\left(\frac{f_0^\beta}{g_0^\beta}\right) = \sgn(E - \beta b_3) .
\end{equation}
We choose to define $f_0^\beta, g_0^\beta$ as positive and place this relative sign in the first two components.

\subsubsection{General Case: $n>0$}
For $n>0$, we can use the explicit forms of $c_n^\sigma$ and $d_n^\sigma$ as well as the dispersion relation to find
\begin{equation}
    \left(\frac{f_n}{g_n}\right)^\sigma = \frac{\sigma \sgn(b_0+\sigma b_3)}{2m} ((\tilde{p}+\sigma b)^2 - m^2) \sqrt{\frac{1}{b^2} \frac{b_0 (E^2 - \Lambda_n^2) - 2 \sigma b_3 X}{b_0 (E^2 - \Lambda_n^2) + 2 \sigma b_3 X}} ,
\end{equation}
where again $\tilde{p}^\mu = (E,0,\sqrt{2n|qB|},p_3)$. Normalizing $\psi_\beta^{(n)}$ gives
\begin{gather}
    |f_n|^2 = \frac{((\tilde{p} + b)^2 - m^2) (b_0 (E^2 - \Lambda_n^2) - 2 b_3 X)}{4 b^2 (E (E^2 - \Lambda_n^2) - 2 b_3 X)} ,
    \label {eq:Fsq} \\
    |g_n|^2 = -\frac{((\tilde{p} - b)^2 - m^2) (b_0 (E^2 - \Lambda_n^2) + 2 b_3 X)}{4 b^2 (E (E^2 - \Lambda_n^2) - 2 b_3 X)} .
    \label{eq:Gsq}
\end{gather}
The minus sign in $|g_n|^2$ may seem alarming, but recall the second form of the dispersion given by Eq. (\ref{eq:Dispersion2}). The relative sign between $f_n$ and $g_n$ is
\begin{equation}
    u := \sgn\left(\frac{f_n}{g_n}\right) = \sgn\left(\frac{(\tilde{p} + \sigma b)^2 - m^2}{b_0 + \sigma b_3}\right)
\end{equation}
for either choice of $\sigma$. As for the $n=0$ case, we insert this factor by hand in the first two components.

\subsubsection{Summary of Dirac Spinors}
Looking at Eqs. $(\ref{eq:Csq}), (\ref{eq:Dsq}), (\ref{eq:Fsq}),$ and $(\ref{eq:Gsq})$, there is a simplification that can be made: namely, the denominators of $c_n^\sigma$ and $d_n^\sigma$ can be canceled with part of the numerators of $f_n$ and $g_n$. At the same time, we can consolidate our notation. Define, for $n>0$,
\begin{gather}
    C_{\sigma \tau} = \sqrt{\frac{-s^\sigma}{2} \frac{E^2 - \Lambda^2 + 2\tau X}{b_0 - \sigma \tau b_3}} , \\
   G_\sigma = \sqrt{\frac{\sigma s^\sigma}{4} \frac{(\tilde{p} - \sigma b)^2 - m^2}{E (E^2 - \Lambda^2) - 2 b_3 X}} .
\end{gather}
These depend on $n$, but we suppress this index for brevity; for our applications there should be no confusion. The signs $-s^\sigma$ and $\sigma s^\sigma$ are to avoid careless squaring; absolute values bars inside the square roots suffice. Then, collecting our results,
\begin{gather}
   \psi_{\beta su} = e^{-iEt + ip_3 z}
    \begin{pmatrix}
        u G_- \begin{pmatrix} C_{-+} \phi_{n-\bar{\beta}} \\ s^- C_{--} \phi_{n-\overline{-\beta}} \end{pmatrix} \\
        G_+ \begin{pmatrix} C_{++} \phi_{n-\bar{\beta}} \\ s^+ C_{+-} \phi_{n-\overline{-\beta}} \end{pmatrix}
    \end{pmatrix} , 
    \label{eq:DiracSpinorFinal} \\
    s^\sigma = - \sgn\left(\frac{E^2 - \Lambda^2 + 2 \tau X}{b_0 - \sigma \tau b_3}\right) = \sigma \sgn\left(\frac{(\tilde{p} - \sigma b)^2 - m^2}{E (E^2 - \Lambda^2) - 2 b_3 X}\right) , \\
    u = \sigma \sgn\left(\frac{(\tilde{p} + \sigma b)^2 - m^2}{b_0 + \sigma b_3}\right) = \sgn\left(\frac{E^2 - \Lambda^2 + 2X}{E (E^2 - \Lambda^2) - 2 b_3 X}\right) , \\
    (E^2 - \Lambda^2)^2 = 4(X^2 + b^2 2n|qB|) .
\end{gather}
We call $u$ and $s=s^+$ the $u-$ and $s-$helicity of $\psi$.

For the $n=0$ case, define
\begin{equation}
    G_{\beta\sigma} = \sqrt{\frac{E - \beta b_3 + \sigma\beta (p_3 - \beta b_0)}{2(E - \beta b_3)}}
\end{equation}
so that
\begin{equation}
    \psi_{\beta}^{(0)} = e^{-iEt + ip_3z}
    \begin{pmatrix}
        \sgn(E - \beta b_3) G_{\beta-} \\
        G_{\beta+}
    \end{pmatrix} \otimes
    \begin{pmatrix} \frac{1+\beta}{2} \phi_{n-\bar{\beta}} \\ \frac{1-\beta}{2} \phi_{n-\overline{-\beta}} \end{pmatrix} .
\end{equation}
When $n = 0$, the dispersion relation is
\begin{equation}
    E^2 - \Lambda^2 - 2 \beta X = 0 ,
\end{equation}
which reduces $C_{\sigma\tau} \to \delta_{\tau \beta}$ and $G_\sigma \to G_{\beta\sigma}$ (after using absolute values to get rid of the $s^\sigma$'s).

We have defined the signs $s^\sigma$ and $u$ by assuming that the solution to the dispersion (\ref{eq:Dispersion1}) is known. We have called these the $u-$ and $s-$helicity above, for good reason. A straightforward computation yields
\begin{equation}
    \begin{split}
        \frac{1}{m} \int d^3x \, \bar{\psi} \Sigma^3 \psi &= \frac{2X}{E(E^2 - \Lambda^2) - 2 b_3 X} \\
        &\stackrel{b_3=0}{=} \frac{s \,p_3}{E \sqrt{p_3^2 + 2n|qB|}} \\
        &\stackrel{b_0=0}{=} \frac{u}{\sqrt{E^2 - 2n|qB|}} \,,
    \end{split}
\end{equation}
showing that in these cases $s$ and $u$ become helicity and spin. In \cite{AlanKostelecky:2024psy}, the authors identified
\begin{equation}
    \mathcal{S} = -2 \gamma^5 (b \cdot i\partial - m \slashed{b})
\end{equation}
as an operator whose eigenvalues distinguish eigenspinors in the case $b_0=0$. Our spinors are indeed eigenspinors of $\mathcal{S}$, with eigenvalue $2\xi \sqrt{(b \cdot \tilde{p})^2 - m^2 b^2}$, where $\xi = \sgn(p^2 + b^2 - m^2)$. In special cases, we find
\begin{equation}
    \begin{split}
        \xi &\stackrel{b_3=0}{=} \sgn\left(b_0^2 - p_3 b_0 \sqrt{p_3^2 + 2n|qB|}\right) \\
        &\stackrel{b_0=0}{=} \sgn(u b_3) \,.
    \end{split}
\end{equation}
However, we have been unable to find a pair of operators which directly gives $u$ and $s$ (or some product thereof) in the general setting.

\subsection{Limiting Cases}
Here we gather the results of setting $b_0$ or $b_3$ to 0. It should be noted that these operations do not commute, for the reasons discussed above. We also include the case $m=0$.

\subsubsection{Chiral chemical potential only: $b_3 = 0$}
In this case the dispersion is
\begin{equation}
    E = u \sqrt{m^2 + (|\bm{p}| - s b_0)^2} ,
    \label{eq:b0Disp}
\end{equation}
where $|\bm{p}| = \sqrt{p_3^2 + 2n|qB|}$ (and $n=0$ only allowing $s = \beta$), and the spinor coefficients are
\begin{equation}
    \begin{gathered}
        G_\sigma = \frac{1}{2} \sqrt{\frac{E - \sigma b_0 + \sigma s |\bm{p}|}{E |\bm{p}|}} , \qquad
        C_{\sigma\tau} = C_\tau = \sqrt{|\bm{p}| + \tau s p_3} , \\
        s^\sigma = s = - \sgn(b_0 (E^2 - \Lambda^2)) , \qquad
        u = \sgn(E) .
    \end{gathered}
\end{equation}
These were also derived by \cite{Sheng:2017lfu}. Apart from $s$, this is well-defined in the subsequent limit $b_0 \to 0$, where $s$ can be replaced with helicity.

\subsubsection{Weyl semimetal: $b_0 = 0$}
In this case the dispersion is
\begin{equation}
    E = \pm \sqrt{2n|qB| + \left(\sqrt{m^2 + p_3^2} + u b_3\right)^2} ,
    \label{eq:b3Disp}
\end{equation}
(with $n=0$ only allowing $u = \beta$) and the spinor coefficients are
\begin{equation}
    \begin{gathered}
        G_\sigma = \frac{1}{2} \sqrt{\frac{\sqrt{m^2 + p_3^2} + \sigma u p_3}{|E| \sqrt{m^2 + p_3^2}}} , \qquad
        C_{\sigma\tau} = C_\tau = \sqrt{|E| + \tau \, \sgn(E) \left(b_3 + u \sqrt{m^2 + p_3^2}\right)} , \\
        s^\sigma = \sigma \sgn(E), \qquad u = \sgn(b_3) \sgn(E^2 - \Lambda^2 - 2b_3^2) .
    \end{gathered}
\end{equation}
These are also well-defined when $b_3 \to 0$, provided $u$ is replaced with spin. Notice that the limits $b_0 \to 0$ and $b_3 \to 0$ do not commute; they recover solutions to the ordinary Dirac equation that are eigenspinors of different operators.

\subsubsection{Massless limit: $m = 0$}
\label{sec:LimitMassless}

In this case the Dirac equation fully decouples the left- and right-chiral parts of the Dirac spinor, leaving $\Pi_\sigma \chi_\sigma = 0$. This simplifies the analysis considerably. The resulting dispersion relation is
\begin{equation}
    E = -\sigma b_0 + r \sqrt{(\sigma p_3 + b_3)^2 + 2n|qB|} ,
\end{equation}
with $r = \sgn(E+\sigma b_0)$. In the special case $n=0$, only $r = \beta$ is allowed. The corresponding eigenspinors (for any $n$) are
\begin{equation}
    \begin{gathered}
        \psi_{r \sigma} = e^{-iEt + i p_3 z}
        \begin{pmatrix}
            \frac{1-\sigma}{2} \chi_- \\
            \frac{1+\sigma}{2} \chi_+
        \end{pmatrix} , \quad
        \chi_\sigma =
        \begin{pmatrix}
            M_{\sigma +} \phi_{n+\frac{1-\beta}{2}} \\
            r \sigma M_{\sigma -} \phi_{n+ \frac{1+\beta}{2}}
        \end{pmatrix} \\
        M_{\sigma \tau} = \sqrt{\frac{E + \sigma b_0 + \tau (\sigma p_3 + b_3)}{2 (E + \sigma b_0)}} .
    \end{gathered}
\end{equation}
It should be noted that in these expressions, the only effect of $b_\mu$ is to shift $p_\mu$ by $\sigma b_\mu$.

\subsection{Choosing a Gauge}
We now choose $a = 1/2$, i.e.\ the symmetric gauge $A_\mu = \frac{1}{2} B (0,y,-x,0)$. This gauge is best treated with cylindrical coordinates $(r,\phi,z)$, though the coordinate $\rho = \frac{|qB|}{2} r^2$ is more useful than $r$. The eigenfunctions $\phi_n$ in this gauge are
\begin{equation}
    \phi_n = \begin{pmatrix} \phi_n^{(1)} \\ \phi_n^{(2)} \end{pmatrix} \sim e^{ij\phi} \begin{pmatrix} e^{-i\phi/2} I_{n-\frac{1-\beta}{2},a} \\ -i\beta e^{i\phi/2} I_{n-\frac{1+\beta}{2},a} \end{pmatrix} ,
\end{equation}
with $j = \beta(a-n+1/2)$ the $z$-angular momentum quantum number and $a$ now a nonnegative integer. The functions $I_{n,a}(\rho)$ are the Laguerre functions
\begin{equation}
    I_{n,a}(\rho) = \sqrt{\frac{a!}{n!}} e^{-\rho/2} \rho^{\frac{n-a}{2}} L_{a}^{(n-a)}(\rho) ,
\end{equation}
with $L_a^{(n-a)}(\rho)$ an associated Laguerre polynomial. In the previous section we ignored the two-component nature of $\phi_n$ as well as the index $j$. Rest assured, these take care of themselves: see Appendix \ref{app:LandauLevels}. The two-component spinors of the previous section are then
\begin{equation}
    \chi_\sigma^{(n)} = e^{ij\phi} \begin{pmatrix} C_{\sigma +} e^{-i\phi/2} I_{n-\frac{1-\beta}{2},a} \\ -i\beta s^\sigma C_{\sigma -} e^{i\phi/2} I_{n-\frac{1+\beta}{2},a} \end{pmatrix} .
\end{equation}
The functional form is identical to the $b_\mu = 0$ case; the only difference is in the coefficients $C_{\sigma \tau}$ and $G_\sigma$. The same would be true if we had chosen the Landau gauge $a = 1$.

\subsection{Chiral magnetic current}

In Appendix~\ref{sec:CME}, we use the solutions to the Dirac equation with finite $b_0$ and $b_3$ (\ref{eq:SpinorNormalized}) that we derived in this section to compute the chiral magnetic current. We verify that the computed current does not depend on $b_3$, as expected.

\section{Synchrotron Radiation}
\label{sec:Rad}
We now move to computing the synchrotron radiation emitted from the fermions described above. The $\mc{S}$-matrix element we wish to compute is
\begin{equation}
    \mc{S} = \int  \bar{\psi}_{s'u'\beta}^{(n')}iq \slashed{A}^* \psi_{su\beta}^{(n)} \, d^4 x \, .
\end{equation}

The Dirac spinors, properly normalized, are
\begin{equation}
    \psi_{su\beta} = e^{-i Et + ip_3 z + ij\phi} \sqrt{\frac{|qB|}{2\pi L}}
    \begin{pmatrix} C_1 e^{-i\phi/2} I_1 \\ -i\beta C_2 e^{i\phi/2} I_2 \\ C_3 e^{-i\phi/2} I_1 \\ -i\beta C_4 e^{i\phi/2} I_2 \end{pmatrix} = e^{-i Et + ip_3 z + ij\phi} \sqrt{\frac{|qB|}{2\pi L}} \chi ,
    \label{eq:SpinorNormalized}
\end{equation}
We have written
\begin{equation}
    I_1(\rho) = I_{n-\frac{1-\beta}{2},a}(\rho), \qquad I_2(\rho) = I_{n-\frac{1+\beta}{2},a}(\rho) ,
\end{equation}
and $C_i$ for the coefficients determined in the previous section (not to be confused with $C_{\sigma\tau}$).

The photon wavefunctions, with helicity $h$, $z$-angular momentum $\ell$, and momentum $\bm{k}$, are
\begin{equation}
    \begin{gathered}
        \bm{A}_{h,\ell,\bm{k}}(x) = \frac{1}{\sqrt{2\omega V}} \Phi_{h,\ell,\bm{k}}(\bm{x}) e^{-i\omega t} , \qquad
        \Phi_{h,\ell,\bm{k}}(\bm{x}) = \frac{1}{\sqrt{2}} (h \bm{T} + \bm{P}) e^{ik_z z + i\ell \phi} , \\
        \bm{T} = \frac{i \ell}{k_\perp r} J_\ell(k_\perp r) \hat{r} - J_\ell'(k_\perp r) \hat{\phi} , \qquad 
        \bm{P} = \frac{i k_z}{k} J_\ell'(k_\perp r) \hat{r} - \frac{\ell k_z}{k k_\perp r} J_\ell(k_\perp r) \hat{\phi} + \frac{k_\perp}{k} J_\ell(k_\perp r) \hat{z} .
    \end{gathered}
\end{equation}

The calculation is the same as in the $b_0=b_3=0$ case (a classic reference is \cite{sokolov1986radiation}; we recently reviewed the calculation, with modern notation, in \cite{Buzzegoli:2023vne}). A straightforward but tedious calculation, whose only non-trivial step is an integral identity \cite{kolbig_hankel_1996}, brings us to
\begin{equation}
    \mc{S} = 2\pi \delta(E-E'-\omega) \frac{2\pi}{L} \delta(p_3-p_3'-k_z) \delta_{l,j-j'} \beta^{j-j'} I_{a,a'}(x) \frac{-iq}{\sqrt{2\omega V}} \langle j \cdot \Phi \rangle,
\end{equation}
with
\begin{equation}
        \sqrt{2} \langle j \cdot \Phi \rangle = -K_1 (h - \cos\theta) I_{n-1,n'} + K_2 (h+\cos\theta) I_{n,n'-1} + \sin(\theta) (K_3 I_{n,n'} - K_4 I_{n-1,n'-1})
\end{equation}
for $\beta = +1$; for $\beta = -1$, switch $n \leftrightarrow n-1$ and $n' \leftrightarrow n'-1$. The coefficients $K_i$ are
\begin{equation}
    \begin{gathered}
        K_1 = C_3' C_4 - C_1' C_2 , \qquad K_2 = C_4' C_3 - C_2' C_1 , \\
        K_3 = C_3' C_3 - C_1' C_1 , \qquad K_4 = C_4' C_4 - C_2' C_2 .
    \end{gathered}
\end{equation}

From here, we compute the differential rate
\begin{equation}
    \frac{d \dot{w}}{d\Gamma d\Gamma' d\Gamma_k} = \frac{|\mc{S}|^2}{T} = \frac{q^2}{2\omega V} 2\pi \delta(E-E'-\omega) \frac{2\pi}{L} \delta(p_3-p_3'-k_z) \delta_{l,j-j'} I_{a,a'}^2 |\langle \bm{j \cdot \Phi} \rangle|^2,
    \label{eq:RatePrimitive}
\end{equation}
with $T$ a formal symbol to cancel a factor of $2\pi \delta(E-E'-\omega)$ (also called the observation time). $d\Gamma = \frac{d^3x d^3k}{(2\pi)^3}$ is a phase space differential for the initial fermion and $d\Gamma'$ and $d\Gamma_k$ are the same for the final fermion and photon. Integrating the photon spatial volume and final fermion phase spaces, we find
\begin{equation}
    \frac{d\dot{w}}{d\Gamma d^3k} = \frac{q^2}{4\pi} \frac{1}{2\pi\omega} \sum_{n',a'} \delta_{l,j-j'} I_{a,a'}(x)^2 |\langle \bm{j \cdot \Phi} \rangle|^2 \delta(E-E'-\omega),
\end{equation}
with $p_3' = p_3 - k_z$, and we have used the substitution
\begin{equation}
    \int \frac{d^3p'}{(2\pi)^3} \to \frac{1}{A} \sum_{n',a'} \int \frac{dp_3'}{2\pi}
\end{equation}
appropriate to the Landau level description \cite{mameda_chiral_2017}. $A$ is the area of the extent of our spatial cylinder, canceled by the integration $\int d^3x' = V$. The same will be done for $d\Gamma$ later. We will use another trick for the photon momentum:
\begin{equation}
    \int d^3 k \enspace \delta_{l,j-j'} = 2\pi \sum_l \delta_{l,m-m'} \int dk_z dk_\perp k_\perp = 2\pi \int dk_z dk_\perp k_\perp = \int d^3k ,
\end{equation}
where we have used the azimuthal symmetry of the problem in the last equality.
Then we can use the identity \cite{sokolov1986radiation}
\begin{equation}
    \sum_{a'} I^2_{a,a'}(x) = 1
\end{equation}
to find the differential photon emission rate: 
\begin{equation}
    \frac{d\dot{w}}{d\Gamma d^3k} = \frac{q^2}{4\pi} \frac{1}{2\pi\omega} \sum_{n'}|\langle \bm{j \cdot \Phi} \rangle|^2 \delta(E-E'-\omega),
    \label{eq:RateReduced}
\end{equation}
evaluated with $p_3' = p_3-k_z$. We have suppressed the potential dependence on the final state's helicity quantum numbers, which are not specified by fixing $p_3'$ and $n'$. The photon helicity dependence is also suppressed, but this dependence is easily computable given the expression for the matrix elements.

\subsection{Single-particle radiation}
Here we evaluate the radiation intensity $W$, the radiated energy per unit time, from a single particle, i.e.
\begin{equation}
    W(E,n,p_3) = \int d^3k \hspace{3pt} \omega\frac{d\dot{w}}{d\Gamma d^3 k}
\end{equation}
for a specified initial state with quantum numbers $E,n,p_3$. Writing out the rate integral,
\begin{equation}
    \dot{w}(E,n,p_3) = \frac{q^2}{4\pi} \sum_{n'} \int d\omega d\cos\theta \hspace{3pt} \omega |\langle \bm{j \cdot \Phi}\rangle|^2 \delta(E-E'-\omega)
\end{equation}
(we have dropped $d\Gamma$ in favor of the functional dependence $(E,n,p_3)$). The Dirac delta can be converted to fix $\omega$ with the factor
\begin{gather}
    \delta(E-E'-\omega) = \frac{\delta(\omega - \omega_0)}{\left| 1 + \frac{\partial E'}{\partial \omega} \right|} = \delta(\omega-\omega_0) |\Delta|, \\
    \Delta = \frac{E'(E'^2 - \Lambda'^2) - 2b_3 X'}{(E'-p_3'\cos\theta)(E'^2 - \Lambda'^2) - 2(b_3 - b_0 \cos\theta) X'} ,
\end{gather}
where $\omega_0$ is the solution to $E = E' + \omega$. Multiple solutions may exist, corresponding to different values of $s^\sigma{}'$ and $u'$. The result is
\begin{equation}
    W(E,n,p_3) = \frac{q^2}{4\pi} \sum_{n'} \int d\cos\theta \hspace{3pt} \omega_0^2 |\langle \bm{j \cdot \Phi}\rangle|^2 |\Delta| .
    \label{eq:SPInt}
\end{equation}
We compute this quantity numerically in Section \ref{sec:NumSP}.

\subsection{Plasma radiation}
To compute the radiation from a plasma in thermal equilibrium, we go back to Eq. (\ref{eq:RatePrimitive}) and apply Fermi-Dirac statistical weight factors. Since the manipulations leading to Eq. (\ref{eq:RateReduced}) are not affected by this, we can proceed from there. We again use the Landau level prescription for the transverse momentum integration. The Dirac delta can be converted to integrate $p_3$ immediately,
\begin{gather}
    \delta(E-E'-\omega) = \frac{\delta(p_3-p_3^0)}{|\frac{\partial E}{\partial p_3} - \frac{\partial E'}{\partial p_3'}|} = \delta(p_3 - p_3^0) |\Delta| , \\
    \Delta = \frac{(E (E^2 - \Lambda^2) - 2 b_3 X) (E' (E'^2 - \Lambda'^2) - 2 b_3 X')}{(p_3 E' - p_3' E) (E^2 - \Lambda^2) (E'^2 - \Lambda'^2) + 2 (\tilde{p} \cdot b \hspace{3pt} X' (E^2 - \Lambda^2) - \tilde{p}' \cdot b \hspace{3pt} X (E'^2 - \Lambda'^2))} ,
\end{gather}
where $p_3^0$ is a solution to the energy-momentum conservation. Then the number of photons emitted from the plasma per unit time per unit volume for given $\bm{k}$ is
\begin{equation}
    \frac{d N}{dtdV d^2k_\perp dy} = \frac{q^2}{4\pi} \frac{1}{(2\pi)^2} \frac{|qB|}{2\pi} \sum_{n,n',p_3^0} |\Delta| f(E) f(-E') |\langle \bm{j \cdot \Phi} \rangle |^2 .
    \label{eq:PlDiffRate}
\end{equation}
We have changed coordinates to those used in heavy-ion collisions and multiplied by $\omega$ in anticipation of the next section. We are only interested in $y=0$, i.e. the plane perpendicular to the collision axis, where $k_T = \omega$ and $\phi = \pi - \theta$. The sum over $p_3^0$ is a sum over solutions to the energy-momentum conservation condition. When we consider quarks, we multiply this number by the number of colors $N_c$.
The integrated quantities we wish to compute are
\begin{gather}
    \left\langle \frac{dN}{d^2k_\perp dy}\Big|_{y=0} \right\rangle = L \Delta t \frac{1}{2\pi} \int_0^{2\pi} d\phi \hspace{3pt} \frac{dN}{dt dz d^2k_\perp dy}\Big|_{y=0} , 
    \label{eq:v0} \\
    v_2 = \frac{1}{\left\langle \tfrac{dN}{d^2k_\perp dy}|_{y=0} \right\rangle} \frac{1}{2\pi} \int_0^{2\pi} d\phi \hspace{3pt} \cos(2\phi) \frac{dN}{d^2k_\perp dy}\Big|_{y=0} .
    \label{eq:v2}
\end{gather}
Using the rotational symmetry about the $\bm{B}$ axis, the $\phi$ integration can be carried out over $(-\pi/2,\pi/2)$, then multiplied by 2. $L$ and $\Delta t$ are the length of the cylinder of plasma and the observation time interval, roughly the longitudinal size and lifetime of the plasma. We compute these quantities numerically in Section \ref{sec:NumPl}

\section{Synchrotron radiation by QGP at finite $b_0$ and $b_3$}
\label{sec:Num}

In this section, we present results from the numerical calculations of synchrotron radiation from a single particle (Eq. (\ref{eq:SPInt})) and for $\left\langle \tfrac{dN}{d^2k_\perp dy}|_{y=0} \right\rangle$ and $v_2$ of a plasma of fermions (Eqs. (\ref{eq:v0}) and (\ref{eq:v2})) in the presence of nonzero chiral chemical potential $b_0$ and chiral gradient $b_3$. The single-particle radiation is somewhat academic, but hints at the results for $\left\langle \tfrac{dN}{d^2k_\perp dy}|_{y=0} \right\rangle$. The plasma radiation is presented alongside data from the PHENIX detector at RHIC \cite{PHENIX:2014nkk,PHENIX:2022rsx,PHENIX:2015igl,PHENIX:2025ejr}.
 
For a single fermion, we use a mass of $m=300$~MeV, corresponding to the constituent quark mass. In QGP, however, the quark masses are close to their current values, of order a few MeV. Interactions with the medium further modify the quark dispersion relation through the thermal self-energy. The resulting thermal mass is not a scalar mass term in the Dirac equation but arises from a momentum-dependent self-energy that preserves chiral symmetry while breaking Lorentz boost invariance due to the preferred rest frame of the plasma. Importantly, this description is derived within the hard-thermal-loop framework and applies to soft fermionic modes with momenta of order $gT$. Such modes provide only a limited contribution to photon production in a strong magnetic field, where harder fermionic modes are expected to dominate.

For hard modes, a common approach is to employ a quasiparticle description in which the medium effects are incorporated through an effective mass parameter, typically chosen to reproduce the asymptotic thermal dispersion relation. This presents a modeling ambiguity. Using a phenomenological effective mass of $m=300$~MeV provides a reasonable estimate of the quasiparticle energy scale but also introduces explicit chiral symmetry breaking that is absent for the thermal mass. On the other hand, employing the current quark mass, $m\simeq 3$~MeV, correctly reflects the small explicit chiral symmetry breaking of QCD, but underestimates the quasiparticle energy and, consequently, modifies the predicted radiation intensity. Since neither approximation is entirely satisfactory, we present calculations for both mass values to assess the sensitivity of our results to this modeling choice. As we will show, increasing the mass enhances the relative effects of $b_0$ and $b_3$, while reducing the overall photon emission rate.

The chiral chemical potential in relativistic heavy-ion collisions can be as large as $100$~MeV \cite{Lin2018AxialCharge,Ghosh:2021naw,Fukushima:2010fe,Kovalenko:2023xpg,Yu:2014sla,Frenklakh:2024jff}. We assume that the chiral gradient $b_3$ may have similar order of magnitude. The numerical calculations presented in this section explore a range of parameters. 

\subsection{Single-particle radiation}
\label{sec:NumSP}

\begin{figure}
    \centering
    \includegraphics[width=0.95\linewidth]{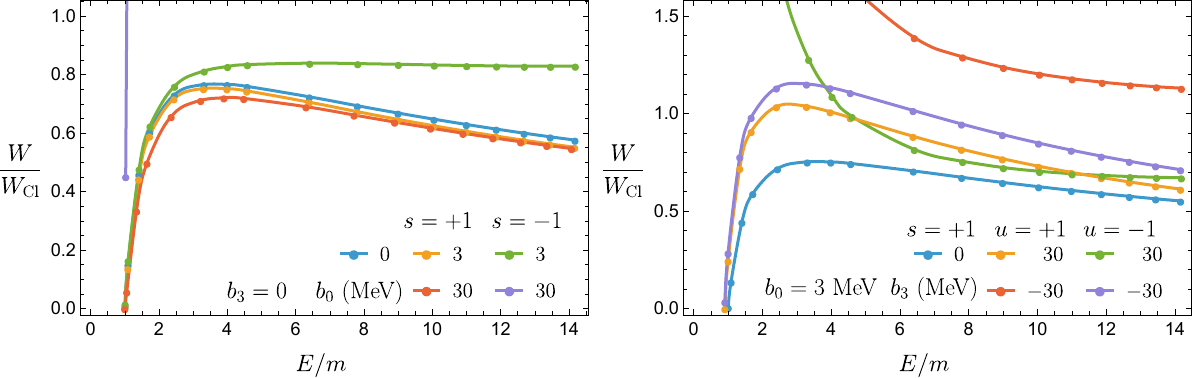}
    \caption{The single-particle radiation intensity due to a magnetic field $qB = 0.01 m^2 = 900$ MeV$^2$. Left: The chiral chemical potential $\mu_5 = b_0$ is varied while $b_3 = 0$ is fixed. Right: $\mu_5 = b_0 = 3$ MeV is fixed while $b_3$ varies. The blue curve in the left figure is the $b_0 = b_3 = 0$ result, well-approximated by standard semiclassical computations \cite{Buzzegoli:2023vne}. Differences in $s-$ and $u-$helicity have been made apparent.}
    \label{fig:SP}
\end{figure}

\begin{figure}
    \centering
    \includegraphics[width=0.95\linewidth]{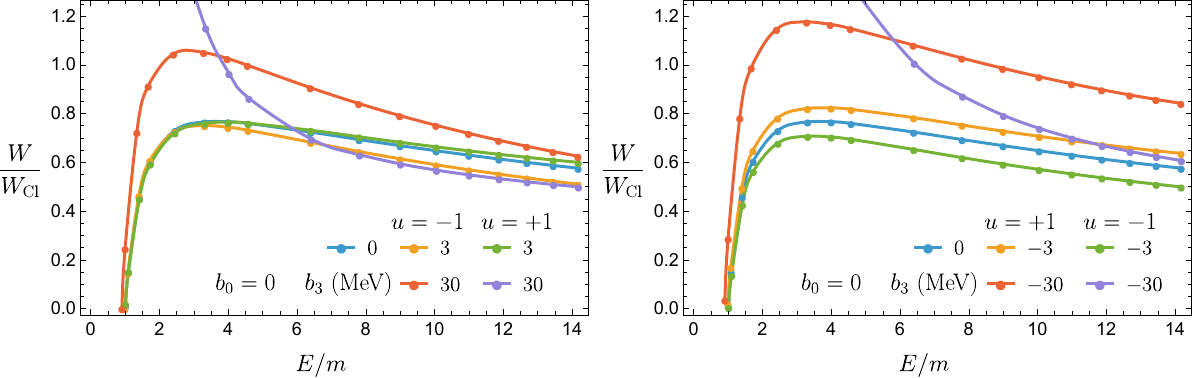}
    \caption{The same as Figure \ref{fig:SP}, but for various values of $b_3$ with $\mu_5 = b_0 = 3$ MeV. The difference due to the sign of $b_3$ is evident.}
    \label{fig:SPb3}
\end{figure}

The calculation of Eq. (\ref{eq:SPInt}) numerically is straightforward. We choose an initial value of $n$ and $p_3$, specifying the momentum of the (single) initial particle, then solve the dispersion relation, Eq.~(\ref{eq:Dispersion1}), numerically.\footnote{The quartic dispersion of course has a solution in terms of radicals, but it is cumbersome to write out, and solving the energy conservation condition for $\omega$ is slower when using these formulae than doing so numerically.} For each $n'$, we numerically solve the energy-momentum conservation condition for the photon energy $\omega$, then compute the matrix element. The sum on $n'$ is finite due to kinematic constraints. $b_0$ and $b_3$ do allow for additional transitions, but their effect is small. We normalize the intensity $W$ to the classical result
\begin{equation}
    W_{\text{Cl}}(E) = \frac{2}{3} \frac{(qB)^2 E^2}{m^4} .
\end{equation}
It should be noted that the value of $B$ used here, as in our previous work \cite{Buzzegoli:2023vne,Kroth:2024mfd}, is not small enough that we expect our results to approach the ratio 1.

Figs.~\ref{fig:SP} and \ref{fig:SPb3} show the single-particle radiation intensity as a function of the initial particle energy $E$ for fixed values of $p_3=0$, $m = 300$~MeV, and $qB = 0.01 m^2 = 900$ MeV$^2$. This value is sufficiently large to facilitate computations for high energies ($< E \approx 15 m = 4500$ MeV) while still being in a regime well-described by semiclassical calculations. We found that changing $p_3$ does not affect any qualitative features of the spectrum.

We note especially the splitting of the radiation from initial particles of opposite $s-$ or $u-$helicity. Recall that these are the parameters which distinguish the energy eigenvalues, and reduce to ordinary helicity and the sign of $E$ when either $b_0$ or $b_3$ is set to 0. The splitting is due to the fact that, even when $b_0 = b_3 = 0$, there is a small difference in radiation when the initial spin is parallel or antiparallel to $q\bm{B}$. Since $s$ and $u$ correspond to spin eigenvalues in this case, this feature is unsurprising. $b_0$ and $b_3$ increase this difference due to the broken degeneracy. When these parameters are larger, the effect is greater. We have neglected to show the full extent of the resulting radiation in the low-energy regime. This enhancement is due to entering a different kinematic regime which is normally accessible by increasing the magnetic field.

The right panel of Fig.~\ref{fig:SP} shows the interplay of the two parameters. Having fixed $b_0= 3$~ MeV, we can see the further effect of including $b_3$ in the calculation. However, there is little that was not covered in the previous discussion.

\subsection{Plasma radiation}
\label{sec:NumPl}

\begin{figure}[t]
    \centering
    \includegraphics[width=\linewidth]{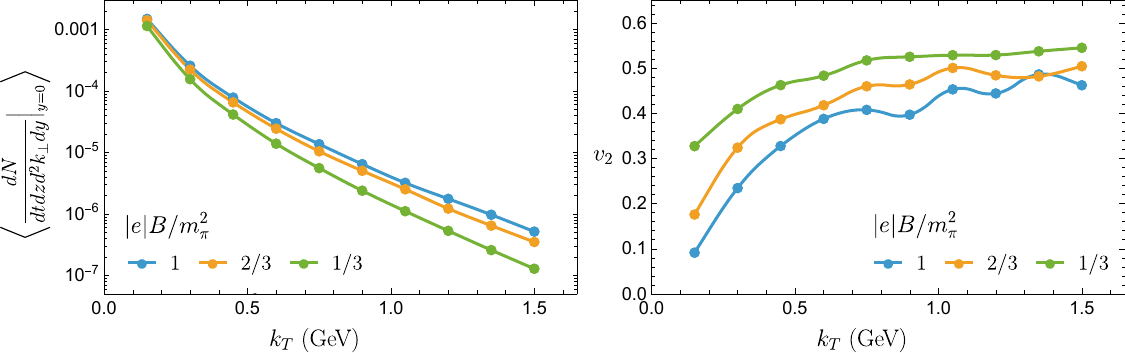}
    \caption{$\left\langle \tfrac{dN}{d^2k_\perp dy}|_{y=0} \right\rangle$ per unit time and length and $v_2$ for the synchrotron photon spectrum of a plasma of fermions with unit charge of mass $m = T = 300$~MeV, three magnetic field strengths, and $b_\mu = 0$.}
    \label{fig:BComp}
\end{figure}

The calculation of $\left\langle \tfrac{dN}{d^2k_\perp dy}|_{y=0} \right\rangle$ and $v_2$ numerically is done in two steps. We first compute the differential rate given by Eq.~(\ref{eq:PlDiffRate}) for fixed $\omega$ as a function of $\theta$, the spherical polar coordinate, then integrate to get either quantity. To compute a summand, a single term in the sum, for given $n,n',\omega$, and $\theta$ we solve the dispersions for the inital and final particle numerically for $E$ and $p_3$ by imposing energy-momentum conservation on the dispersion for $E'$.\footnote{Even when $b_0$ or $b_3$ are 0, and hence $E$ can be written succintly, it is not trivial to write down $p_3^0$, so we still do this step numerically in these cases.} Then the summand can be computed.

\begin{figure}[t]
    \centering
    \includegraphics[width=\linewidth]{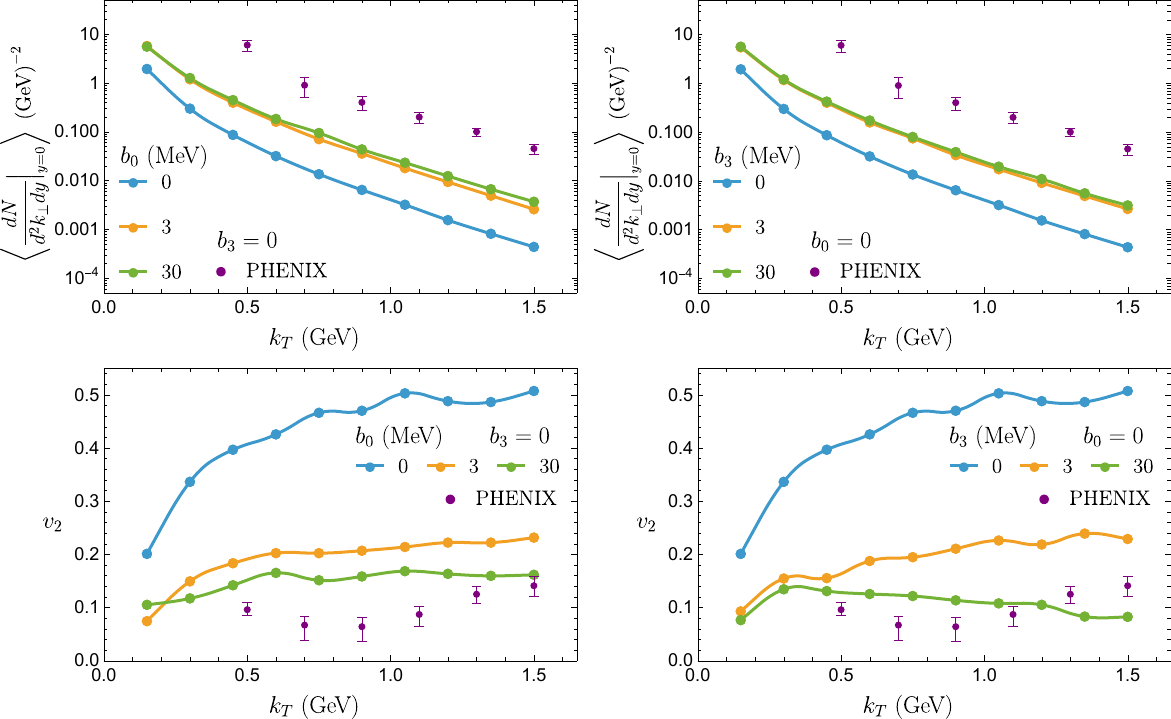}
    \caption{$\left\langle \tfrac{dN}{d^2k_\perp dy}|_{y=0} \right\rangle$ and $v_2$ for the photon spectrum of a plasma of up and down quarks with temperature $T = m = 300$~MeV. $b_0 = \mu_5$ is a chiral chemical potential. The magnetic field is $|e|B = m_\pi^2 = 18000$ MeV$^2$. We have estimated $L = c \Delta t = 10$ fm. The purple points are 20-40\% centrality results from the PHENIX collaboration \cite{PHENIX:2014nkk,PHENIX:2022rsx,PHENIX:2015igl,PHENIX:2025ejr}.}
    \label{fig:PlasmaUpDown}
\end{figure}

\begin{figure}[t]
    \centering
    \includegraphics[width=\linewidth]{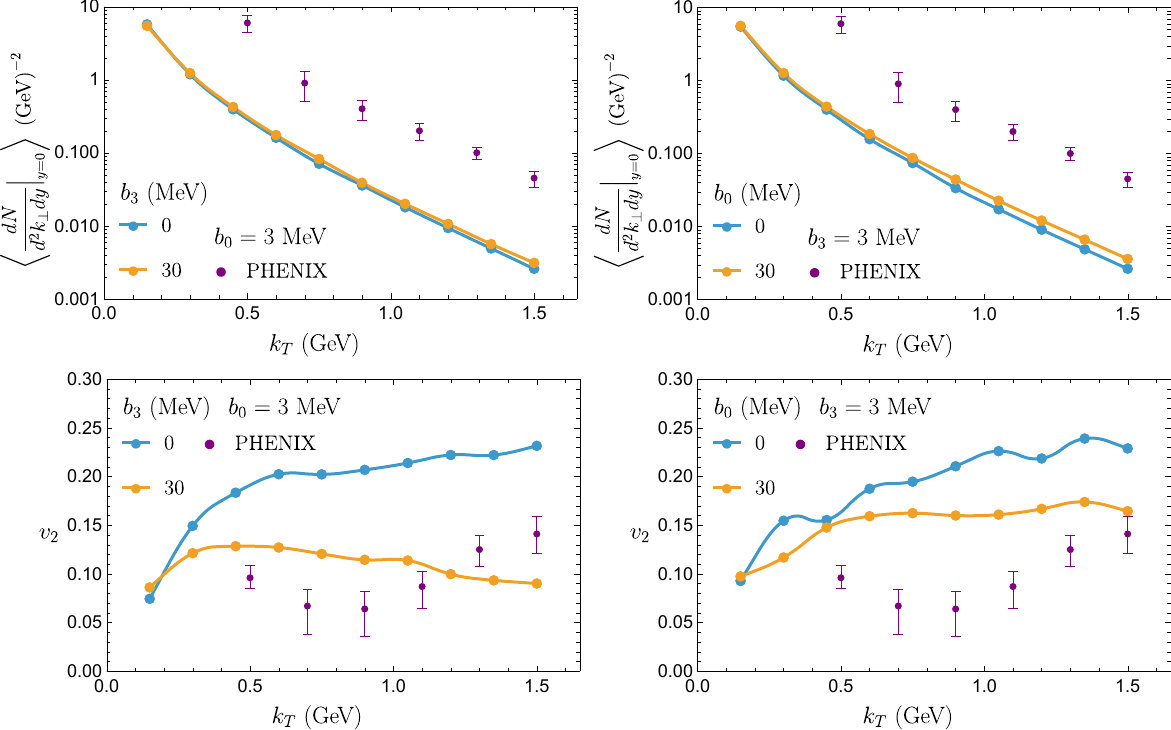}
    \caption{The same as Fig. \ref{fig:PlasmaUpDown}, but with both $b_0 = \mu_5$ and $b_3$ turned on. Left: $b_0 = 3$~MeV is fixed. Right: $b_3 = 3$~MeV is fixed. We have estimated $L = c \Delta t = 10$ fm. The purple points are 20-40\% centrality results from the PHENIX collaboration \cite{PHENIX:2014nkk,PHENIX:2022rsx,PHENIX:2015igl,PHENIX:2025ejr}.}
    \label{fig:PlasmaVaryBoth}
\end{figure}

Since the limits of the sum over $n$ is in principal infinite, we have to cut it off. We do this by summing over 100 initial-state Landau levels at a time, then checking to see if the sum has converged. We stop when the next 100 levels gives less than a 1\% difference in $\left\langle \tfrac{dN}{d^2k_\perp dy}|_{y=0} \right\rangle$. In practice, this required no more than the first 1000 initial-state Landau levels for the magnetic field strengths used. The sum over $n'$ for given $n$ terminates, as in the single-particle case. We again include the additional transitions allowed by $b_0$ and $b_3$, but their effect is small, and they do not contribute any qualitative features to the spectrum of the thermal plasma.

In Fig. \ref{fig:BComp}, we have plotted $\left\langle \tfrac{dN}{d^2k_\perp dy}|_{y=0} \right\rangle$ per unit length and time and $v_2$ for three values of $|e|B$ with $b_0=b_3=0$ as a baseline. Both quantities vary slowly with the field strength, at least for the values shown. Figure \ref{fig:PlasmaUpDown} shows $\left\langle \tfrac{dN}{d^2k_\perp dy}|_{y=0} \right\rangle$ and $v_2$ for a plasma of up and down quarks and antiquarks in a magnetic field of $|e|B = m_\pi^2 = 18000$ MeV$^2$. We have used $m = T = 300$ MeV as our mass scale. In computing $\left\langle \tfrac{dN}{d^2k_\perp dy}|_{y=0} \right\rangle$, we had to choose a length and time scale to compare with experimental data; we chose $L = c \Delta t = 10$ fm, about the length and lifetime of the QGP. The effect of either parameter on $\left\langle \tfrac{dN}{d^2k_\perp dy}|_{y=0} \right\rangle$ is an increase in the number of photons produced by a factor of about 4, depending on $p_T$. The effect seems to saturate when either parameter is on the order $10^{-2} m = 3$~MeV. Furthermore, as Figure \ref{fig:PlasmaVaryBoth} shows, including both parameters gives little appreciable difference in photons produced when compared to including just one. The striking result is in $v_2$, where a significant decrease is observed when either parameter is turned on. Turning on both parameters only seems to allow one parameter to dominate, although this may be due to the range of parameters we have explored. The result for $v_2$ is quite promising. First, it does not depend on the length and time scale multiplying $\left\langle \tfrac{dN}{d^2k_\perp dy}|_{y=0} \right\rangle$. Also, traditional hydrodynamics simulations tend to compute $v_2$ to be smaller than what is observed in experiments \cite{Wu:2025iix,Chatterjee:2013naa,Linnyk:2013wma}. Could the two results meet in the middle?

\begin{figure}[t]
    \centering
    \includegraphics[width=\linewidth]{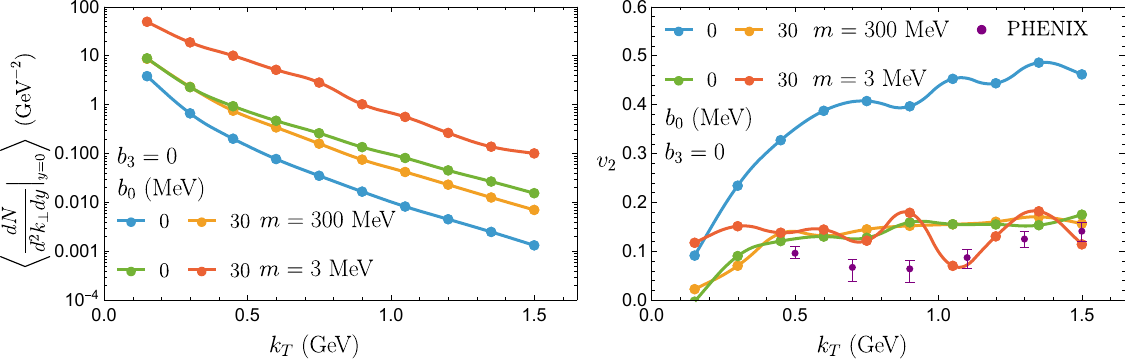}
    \caption{Dependence of the photon spectrum on quark mass. The magnetic field is $|q|B = m_\pi^2 = 18000$~MeV${}^2$, and we have estimated $L = c \Delta t = 10$~fm. The purple points are 20-40\% centrality results from the PHENIX collaboration \cite{PHENIX:2015igl,PHENIX:2025ejr}.}
    \label{fig:MassComp}
\end{figure}

We conclude this section with a discussion of the mass dependence of the radiation from a plasma. In Fig. \ref{fig:MassComp}, we have plotted the photon yield and $v_2$ from unit-charge fermions of mass $m=3$ and $300$~MeV. We see that decreasing $m$ increases the photon yield, but does not otherwise affect the effect of $b_0 = \mu_5$. On the other hand, $v_2$ is significantly decreased when $m$ is decreased, causing $b_0 = \mu_5$ to have little effect on the flow coefficient. This considers the two mass extremes of what we might expect from the QGP, with a more realistic estimate lying somewhere between.

\section{Conclusion}
\label{sec:Conc}

We have solved the modified Dirac equation (\ref{eq:Dirac}) in the presence of a constant magnetic field, a chiral chemical potential $b_0$, and a chirality gradient $b_3$ in the direction of the magnetic field. The solution is presented in Eq.~(\ref{eq:DiracSpinorFinal}). The novel result is the use of $b_3$. We applied this solution to QGP phenomenology by computing synchrotron radiation from fermions in a medium with these properties and found that while the total number of photons $\left\langle \tfrac{dN}{d^2k_\perp dy}|_{y=0} \right\rangle$ produced is only modestly enhanced, the traditionally large synchrotron $v_2$ is significantly reduced to a level comparable to experimental data.

While the decrease in the synchrotron $v_2$ is promising, we should temper our conclusions. As shown in Fig.~\ref{fig:PlasmaUpDown}, the photon yield $\left\langle \tfrac{dN}{d^2k_\perp dy}|_{y=0} \right\rangle$ is still about an order of magnitude smaller than the yield reported by experiments, and this is assuming that the QGP maintains a magnetic field $\sim m_\pi^2$ for the better part of its lifetime. As such, the weight in the $v_2$ contribution from this source will be much smaller than the weight from the more sophisticated hydrodynamic models, such as \cite{Gale:2021emg}. We hope that future hydrodynamic models will begin to include $P$- or $CP$-violating effects to test whether or not they can contribute to a solution to the missing photons puzzle (some work has already been done on this subject \cite{Son:2009tf}, but phenomenological work directed at this problem remains an open problem).

The role of the fermion mass remains subject to a modeling uncertainty in a thermal medium, where the dynamically generated thermal mass differs fundamentally from a Dirac mass by preserving chiral symmetry and modifying the dispersion relation through the medium-induced self-energy. We therefore compared calculations using both current and constituent quark masses, finding that a larger mass enhances the effects of $b_0$ and $b_3$ while suppressing the overall photon emission intensity.

We have demonstrated in our recent publications that plasma rotation increases the number of photons produced \cite{Buzzegoli:2025qfl,Kroth:2024mfd,Buzzegoli:2023vne,Buzzegoli:2022dhw}. Therefore, it is reasonable to expect that the synchrotron radiation from a rotating plasma in the presence of a chiral imbalance could solve the direct photon puzzle. Such a calculation is not prohibitive: taking rotation into account only requires adding $\bm{\Omega} \cdot \bm{J}$ to the Hamiltonian. If the rotation axis is chosen to coincide with $\bm{B}$ (which is on average true in the QGP), this operator commutes with the Hamiltonian associated with the Dirac equation we solved in this paper. The only difference, then, is the dependence of $E$ on the angular momentum quantum number $j$, which only affects the kinematics \cite{Buzzegoli:2023vne}. It should be noted that this introduces certain numerical difficulties and the possibility of requiring new boundary conditions, but such problems have been dealt with before, both exactly \cite{Buzzegoli:2023yut} and with semiclassical methods \cite{Buzzegoli:2025qfl}. Future phenomenological applications may also consider any possible changes in the photon spectrum due to plasma expansion \cite{Gale:2014dfa,Rapp:2016xzw}.

Wang and Shovkovy \cite{Wang:2024gnh} recently published a similar calculation of synchrotron radiation emitted from a plasma of massless fermions with a chiral chemical potential. They were primarily interested in the differences in the spectra of different photon polarizations and did not compute $\left\langle \tfrac{dN}{d^2k_\perp dy}|_{y=0} \right\rangle$ or $v_2$. We have not directly compared to their results because in the massless limit, we find that the matrix elements for the case with $b_0 = \mu_5$ (and $b_3$) are identical to those for the case without them. Because the Dirac equation decouples, the only effect on the wavefunctions is to shift $p_\mu$ by $b_\mu$. Since photon emission from massless fermions cannot induce a chirality change, the kinematics of the reaction is not affected by this shift. As a result, we find that the only difference in the photon spectrum comes from the Fermi-Dirac distribution factors in Eq. (\ref{eq:PlDiffRate}). In this context, our inclusion of a fermion mass makes our work a novel result.

We investigated the role of the chiral transport coefficients $b_0$ and $b_3$ on photon production. It would be interesting to extend this study to include the transverse parameters $b_1$ and $b_2$. This entails solving the Dirac equation for an arbitrary $b_\mu$, a noble goal that requires a dedicated study.

\acknowledgments
We  thank Matteo Buzzegoli and Nandagopal Vijayakumar for many fruitful discussions and collaboration on related projects.  
This work  was supported in part by the U.S. Department of Energy under Grant No.\ DE-SC0023692.

\begin{appendix}
\section{Chiral Magnetic Effect}\label{sec:CME}

Our results allow us to investigate the effect of $b_3$ on the chiral magnetic effect (CME). The CME current can be obtained by computing the velocity expectation value of the wavefunctions in Eq. (\ref{eq:DiracSpinorFinal}). A derivation in the same spirit was present in the original paper of Fukushima et al. \cite{Fukushima:2008xe} using formal properties of the wavefunctions. Our analysis reduces this argument to a straightforward matrix multiplication.

The velocity expectation value is
\begin{equation}
    \bm{v} = \int d^3x \, \bar{\psi}_{rs} \bm{\gamma} \psi_{rs} .
\end{equation}
Inserting the wavefunction in Eq. (\ref{eq:SpinorNormalized}) and introducing the notation
\begin{equation}
    \tilde{\chi}_\sigma = \begin{pmatrix} C_{\sigma +} I_1 \\ C_{\sigma -} I_2 \end{pmatrix} ,
\end{equation}
one finds
\begin{equation}
    \bm{v} = \frac{|qB|}{2\pi} \int d^2 x_\perp \sum_{\sigma = \pm 1} \sigma G_{\sigma}^2 \tilde{\chi}_\sigma^\dagger ( \sigma^1 \hat{r} + \sigma^2 \hat{\phi} + \sigma^3 \hat{z} ) \tilde{\chi}_\sigma \,.
\end{equation}
The orthonormality of the functions $I_{p,q}(x)$ makes the $\hat{r}$ and $\hat{\phi}$ components vanish and the $\hat{z}$-component integration trivial, leaving
\begin{equation}
    \bm{v} = v \hat{z} = \sum_{\sigma,\tau = \pm 1} \sigma \tau G_\sigma^2 C_{\sigma\tau}^2 \enspace \hat{z} .
\end{equation}
Evaluating the sum, one finds
\begin{equation}
    v = \frac{p_3 (E^2 - \Lambda^2) - 2 b_0 X}{E (E^2 - \Lambda^2) - 2 b_3 X} = \frac{\partial E}{\partial p_3},
\end{equation}
using the notation of Eq. (\ref{eq:LambdaX}).

The result $v = \frac{\partial E}{\partial p_3}$ is hardly surprising, and shows that the derivation of $v$ (although not necessarily $\bm{v}$) requires only the dispersion relation, namely Eq. (\ref{eq:Dispersion1}). Then one may take a thermal average,
\begin{equation}
    \langle v \rangle = \sum_{n,a} \sum_{E_i} \int \frac{dp_3}{2\pi} \frac{\partial E_i}{\partial p_3} n_F(E_i) ,
\end{equation}
with $E_i$ the solutions to the dispersion relation for given $n$. We allow negative energies, as when $b_0$ and $b_3$ are non-zero, three solutions to the dispersion relation can be positive (or negative). The sum over $a$ gives the Landau level degeneracy factor $\frac{|qB|}{2\pi}$, and the integration of $p_3$ gives
\begin{equation}
    \langle v \rangle = \frac{|qB|}{2\pi} \sum_{n} \sum_{E_i} -T \lim_{\Lambda \to \infty} \log\left(\frac{1 + \exp(-E_i(\Lambda)/T)}{1 + \exp(-E_i(-\Lambda)/T)} \right) .
    \label{eq:CMESumLimit}
\end{equation}
The CME current $J$ is this quantity times the charge $q$.

The ordinary setting for the CME has $b_3 = 0$. In this case, the energies $E_i$ are
\begin{equation}
    E_{rs} = r \sqrt{m^2 + \left(\sqrt{2n|qB| + p_3^2} - sb_0 \right)^2}
\end{equation}
for $n > 0$. Evidently, $E_{rs}(\Lambda) = E_{rs}(-\Lambda)$, so all limits for $n > 0$ vanish. When $n = 0$,
\begin{equation}
    E_r = r \sqrt{m^2 + (p_3^2 - \beta b_0)^2} ,
\end{equation}
which does not have the previous symmetry. Instead,
\begin{equation}
    \sum_r \lim_{\Lambda \to \infty} \log\left(\frac{1 + \exp(-E_r(\Lambda)/T)}{1 + \exp(-E_r(-\Lambda)/T)} \right) = \frac{1}{T} \lim_{\Lambda \to \infty} E_+(\Lambda) - E_+(-\Lambda) = - \frac{2\beta b_0}{T} ,
\end{equation}
which gives the familiar expression
\begin{equation}
    J = \frac{q^2B}{2\pi^2} b_0
\end{equation}
for the CME current.

We now investigate whether $b_3$ affects the CME current. We will build up to the massive case slowly. In the setting $b_0 = 0$ and $m \neq 0$, the dispersion
\begin{equation}
    E_{ru} = r \sqrt{2n|qB| + \left(\sqrt{m^2 + p_3^2} + u b_3 \right)^2}
\end{equation}
has a $p_3 \to -p_3$ symmetry even when $n = 0$, so the CME current vanishes. If we allow $b_0 \neq 0$ again, but set $m = 0$, the dispersion
\begin{equation}
    E_{r\sigma} = -\sigma b_0 + r \sqrt{(\sigma p_3 + b_3)^2 + 2n|qB|}
\end{equation}
has the symmetry $E_{\sigma r}(p_3) = - E_{-\sigma -r}(-p_3)$ for $n > 0$, so all of these terms vanish in the sum. For $n=0$,
\begin{equation}
    E_\sigma = -\sigma b_0 + \beta (\sigma p_3 + b_3) .
\end{equation}
The sum is more involved in this case, but one finds that all contributions from $b_3$ cancel, and
\begin{equation}
    J = \frac{q^2 B}{2\pi^2} b_0 \,.
\end{equation}

With $b_0, b_3,$ and $m$ all nonzero, the lack of a compact solution $E = \cdots$ to the dispersion relation presents a difficulty. However, we are only interested in the limit $|p_3| \to \infty$, so an asymptotic form will suffice. A tedious analysis shows
\begin{equation}
    E_{r\tau} = r \left(p_3 + \frac{m^2 + 2n|qB|}{2 p_3}\right) - \tau (b_0 - r b_3)
\end{equation}
for $n > 0$. Putting this into Eq. (\ref{eq:CMESumLimit}), we find that the $n>0$ levels contribute $0$ to the CME current. The $n=0$ case is much more straightforward, as
\begin{equation}
    E_r = \beta b_3 + r \sqrt{(p_3 - \beta b_0)^2 + m^2} ,
\end{equation}
which goes through in almost the same way as the $b_3 = 0$ case to yield
\begin{equation}
    J = \frac{q^2 B}{2\pi^2} b_0 .
\end{equation}
Thus $b_3$ does not affect the CME. This result is hardly surprising, given that it is known that $\bm{b}$ induces a current $\bm{b} \times \bm{E}$, and we have no electric field in this setting. The calculation is nonetheless both instructive and reassuring.

\section{Landau Level Wavefunctions}
\label{app:LandauLevels}

In this appendix, we show a few details concerning the spatial eigenfunctions $\phi_n(\bm{x}_\perp)$ used in the main text.

The first item is the fact that $D_\pm^\perp = iD_1 \mp D_2$ serve as raising and lowering operators for the functions $\phi_n$ in a constant magnetic field $\bm{B} = B \hat{z}$. This can be seen from the fact that
\begin{equation}
    [D_+^\perp,D_-^\perp] = 2qB \,,
\end{equation}
which, after relabeling
\begin{equation}
    \hat{a}_\beta = \frac{-1}{\sqrt{2|qB|}} (iD_1 - \beta D_2) = \frac{-1}{\sqrt{2|qB|}} D_\beta^\perp, \quad \hat{a}_\beta^\dagger = \frac{-1}{\sqrt{2|qB|}} (iD_1 + \beta D_2) = \frac{-1}{\sqrt{2|qB|}} D_{-\beta}^\perp ,
\end{equation}
yields $[\hat{a}_\beta,\hat{a}_\beta^\dagger] = 1$. This signals the algebra of a simple harmonic oscillator.

To write a solution to the Dirac equation, we are forced to choose a gauge: we choose the symmetric gauge, in which cylindrical coordinates $(r,\phi,z)$ are most useful. In this gauge,
\begin{equation}
    \hat{a}_\beta = i e^{i\beta \phi} \frac{\sqrt{\rho}}{2} \left(2 \partial_\rho + \frac{i\beta}{\rho} \partial_\phi + 1 \right) , \quad \hat{a}_\beta^\dagger = i e^{-i\beta \phi} \frac{\sqrt{\rho}}{2} \left(2 \partial_\rho - \frac{i\beta}{\rho} \partial_\phi - 1\right) ,
\end{equation}
where $\rho = |qB|r^2/2$. Since we seek spinor solutions, things are a bit trickier. Here we only sketch the solutions, but we recently gave more detail in \cite{Buzzegoli:2023vne}. Since $\gamma^0\gamma^5\slashed{b}$ (the term in the Hamiltonian corresponding to (\ref{eq:Dirac})) commutes with $J_z$, we expect
\begin{equation}
    \psi_n(\bm{x}_\perp) = e^{ij\phi} \begin{pmatrix} e^{-i\phi/2} f_1(r) \\ e^{i\phi/2} f_2(r) \\ e^{-i\phi/2} f_3(r) \\ e^{i\phi/2} f_4(r) \end{pmatrix} = e^{im\phi} e^{-i \Sigma^3 \phi/2} \bm{f} .
\end{equation}
The resulting differential equation for the $f_i$ shows that up to the constants derived in the main text, $f_1(r) = f_3(r)$ and $f_2(r) = f_4(r)$. These solve the differential equations 
\begin{equation}
    \hat{a}_\beta^\dagger \hat{a}_\beta f(r) = - 2|qB| f(r) , \quad \hat{a}_\beta \hat{a}_\beta^\dagger f(r) = -2|qB| f(r) ,
\end{equation}
(which functions solve which equation depends on $\beta$), whose solutions are the Laguerre functions
\begin{equation}
    I_{n,a}(\rho) = \sqrt{\frac{a!}{n!}} e^{-\rho/2} \rho^{(n-a)/2} L_a^{(n-a)}(\rho) ,
\end{equation}
with $a = \beta j + n -1/2$ and $L_a^{(n-a)}(\rho)$ a generalized Laguerre polynomial. The recurrence relations of the Laguerre polynomials imply that
\begin{equation}
    D_\tau^\perp e^{i(j-\tau/2)\phi} I_{n-(1-\tau\beta)/2,a}(\rho) = i\tau\beta\sqrt{2n|qB|} e^{i(j+\tau/2)\phi} I_{n-(1+\tau\beta)/2,a}(\rho) .
    \label{eq:RaiseLowerAppendix}
\end{equation}

This is nearly the statement of Eq. (\ref{eq:RaiseLower}), once we assemble our results properly. The spatial eigenspinors are
\begin{equation}
    \phi_n^\sigma(\bm{x}_\perp) = e^{ij\phi} e^{-i\sigma^3 \phi/2} \begin{pmatrix} c_n^\sigma I_{n-\frac{1-\beta}{2},a}(\rho) \\ -i\beta d_n^\sigma I_{n-\frac{1+\beta}{2},a}(\rho) \end{pmatrix} .
\end{equation}
The coefficients $c_n^\sigma$ and $d_n^\sigma$ are those derived in the text. The full wavefunction is $\psi_n = (f_n \phi_n^- \enspace g_n \phi_n^+)^T$.

The second item is that we did not use the spinor nature of $\phi_n^\sigma$ to find the coefficients of the Dirac spinor. The main jump was getting from Eq. (\ref{eq:DiracSquared}) to to Eq. (\ref{eq:CoeffMatrixEqn}). For simplicity, set $\beta=-1$, and look only at the first row of Eq. (\ref{eq:DiracSquared}), applied to the expansion in the eigenspinors $\phi_n^\sigma$. Eq. (\ref{eq:RaiseLowerAppendix}) shows that
\begin{equation}
    D_-^\perp D_+^\perp \phi_n^{\sigma (1)} = 2n|qB| c_n^\sigma e^{i(j-1/2)\phi} I_{n,a}(\rho) ,
\end{equation}
and that
\begin{equation}
    D_-^\perp \phi_n^{\sigma (2)} = D_-^\perp i d_n^\sigma e^{i(j+1/2)\phi} I_{n,a} = - \sqrt{2n|qB|} d_n^\sigma e^{i(j-1/2)\phi} I_{n-1,a} = -\sqrt{2n|qB|} \frac{d_n^\sigma}{c_n^\sigma} \phi_n^{\sigma(1)} .
\end{equation}
The second row of Eq. (\ref{eq:DiracSquared}) proceeds similarly. The indices are dizzying, but this is exactly what leads to Eq. (\ref{eq:CoeffMatrixEqn}).

\end{appendix}

\bibliography{anom-biblio.bib}

\end{document}